\documentclass[epj]{webofc}
\usepackage[varg]{txfonts}   % Web of Conferences font
\usepackage{hyperref}

\renewcommand*{\eqref}[1]{Eq.~(\ref{eq:#1})}

\newcommand*{\figref}[1]{Fig.~(\ref{fig:#1})}
\newcommand*{\figlab}[1]{\label{fig:#1}}

\usepackage{color}

\wocname{ARENA-2016}
\woctitle{ARENA-2016}
\begin{document}
\title{Modeling of radio emission from a particle cascade in a magnetic field and its experimental validation}

\author{\firstname{Anne} \lastname{Zilles}\inst{1}\fnsep\thanks{\email{anne.zilles@kit.edu}} for the SLAC T-510 collaboration }

\institute{Institut f\"u{}r Experimentelle Kernphysik, Karlsruher Institut f\"u{}r Technologie, 76128 Karlsruhe, Germany}
% \and
%            the second here
% \and
%            Last address
%           }

\abstract{%
The SLAC T-510 experiment was designed to compare controlled laboratory measurements of radio emission of particle showers
to predictions using particle-level simulations, which are relied upon in ultra-high-energy cosmic-ray air shower detection.
Established formalisms for the simulation of radio emission physics, the “endpoint” formalism and the “ZHS” formalism, lead to results which can be explained by a superposition of magnetically induced transverse current radiation and charge-excess radiation due to the Askaryan effect. Here, we present the results of Geant4 simulations for the SLAC T-510 experiment, taking into account the details of the experimental setup (beam energy, target geometry and material, magnetic field configuration, and refraction effects) and their comparison to measured data with respect to e.g. signal polarisation, linearity with magnetic
field, and angular distribution. We find that the microscopic calculations reproduce the measurements within
uncertainties and describe the data well.
}
\maketitle
\section{Introduction}
\label{intro}
\vspace{-0.15cm}
The interpretation of measured data from air-shower events caused by ultra-high energy cosmic rays is based on state-of-the-art Monte-Carlo simulations. Codes like AIRES~\cite{AIRES} and CORSIKA~\cite{CORSIKA} are available for the simulation of the development of extensive air-showers. 
Both tools are based on different hadronic and electromagnetic interaction codes to handle particle interaction for air shower simulations. They have been extended with formalisms to calculate the radio emission by the particle showers.
The two microscopic formalisms calculate the electric field of the moving charged particles in the shower while making no assumptions on the emission mechanisms.
As the basis for the calculation, they use the positions of the simulated particle tracks and their corresponding times, which are available by design of the particle shower simulation codes.
The first approach is called the endpoint formalism~\cite{Endpoint}\cite{REAS3} and is applied in CoREAS~\cite{CoREAS}, a plug-in for CORSIKA. The second, the ZHS formalism~\cite{ZHS}, is applied in ZHAireS~\cite{ZHAireS}, an extension of AIRES. 

For the interpretation of the measured radio signals of extensive air showers, 
the uncertainty in the simulations has to be reduced by validating them in a controlled laboratory environment.  The T-510 experiment at the SLAC National Accelerator Laboratory was performed to verify simulations for the radio emission from particle showers by a comparison to data. 
An overview of the experimental setup of the T-510 experiment as well as the physics motivation are described in~\cite{Katie}. Details on the simulation study can be found in~\cite{AnneICRC}.

\section{Comparison of simulations and data and their interpretation}
\vspace{-0.15cm}
As already published in \cite{PRL}, the time-domain peaks of the magnetic component for the ZHS and endpoints formalisms agree within 3\% while the time-domain peaks of the measured data exceed the simulations by 35\%. The difference could be shown to be of about the same magnitude as the systematic uncertainty due to the first-order reflection at the bottom surface of the target which we discuss in more detail. We found that uncertainties in the index of refraction of the RF absorbing blanket underneath the target dominate the systematic uncertainty.

\subsection{Impact of internal reflections on signal amplitude}
\vspace{-0.15cm}
%%% wie reflection gemacht
Since the frequency-dependent reflection properties of the RF absorbing blanket, on which the target was positioned, are not known, a quantitative modeling of the reflection in the simulation is not possible. 
Therefore, total reflection at the bottom surface with a reflection coefficient of $R=1$  and a refractive index of the blanket larger than the one of the target is assumed. 
Given by the geometry of the target, the reflected signal arrives at the antenna about $1\,\mbox{ns}$ later than the direct signal. Therefore, the direct and the reflected signals overlap and cannot be separated in time. 
A signal which is first reflected at the top surface and then at the bottom will have a time shift of about $7\,\mbox{ns}$ and with a much lower amplitude.

To predict the polarity of the reflection, the signal can be split up into a horizontal and a vertical component.
Due to the geometry, the polarity of the vertical component is inverted for the part of the signal arriving first at the bottom target surface compared to the one directly propagating to the antenna. Here, the horizontal component retains the same polarity as the direct signal, but will experience a sign flip at the boundary to the blanket due to the higher refractive index of the blanket. To predict the impact of the reflection on the radio signal measured at a particular antenna position, a time-delayed signal with equal amplitude but inverted polarity for both components is added to the simulated direct signal. 
%%% Comparison with reflected in 
This was done for the antenna position on the Cherenkov cone at a height of $6.52\,\mbox{m}$ above the beam line (see \figref{withRefl_Cone}). 
A comparison of the measured radio signal in the time domain with the results from the simulations shows that with the inclusion of the reflected signal as explained above a good description of the measured peak amplitude by the simulations is achieved. 
However, the after-pulse is now predicted to be larger than measured.
In addition, a comparison of the frequency spectra leads to the conclusion that the assumption in the modeling of the total reflection for all frequencies is too simple since the simulations including the reflections show a strong suppression of the signal for high frequencies.
\begin{figure}
\centering
% \sidecaption
\includegraphics[width=0.9\textwidth]{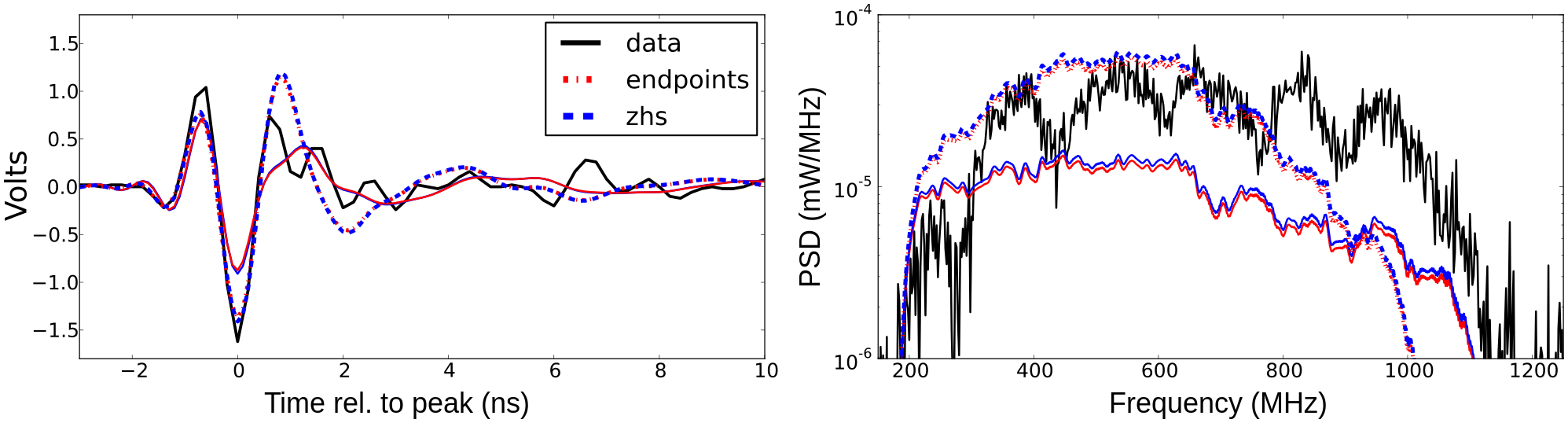}
\caption{Comparison of the measured radio signal at a height of $6.52\,\mbox{m}$ with simulations including the internal reflection (dashed lines) as described in section 2.1. The thin solid lines represent the traces of the simulated results for the direct signal for comparison. Left: Signals in the time domain. Right: Signals in the frequency domain.}
\figlab{withRefl_Cone}       % Give a unique label
\end{figure}
\subsection{Scaling with magnetic field strength}
\vspace{-0.15cm}
To check for the expected linear scaling of the electric field strength of the magnetic component on the magnetic field strength, measurements with different magnetic field strengths were performed~\cite{PRL}. In \figref{Scaling}a the results are compared with those from simulations that include internal reflections. The amplitude of the horizontally polarised component was observed to depend on the magnetic field strength. This supports the conclusion that the transverse current generates the magnetic emission in particle showers.
\begin{figure}
\centering
% \sidecaption
\includegraphics[width=0.9\textwidth]{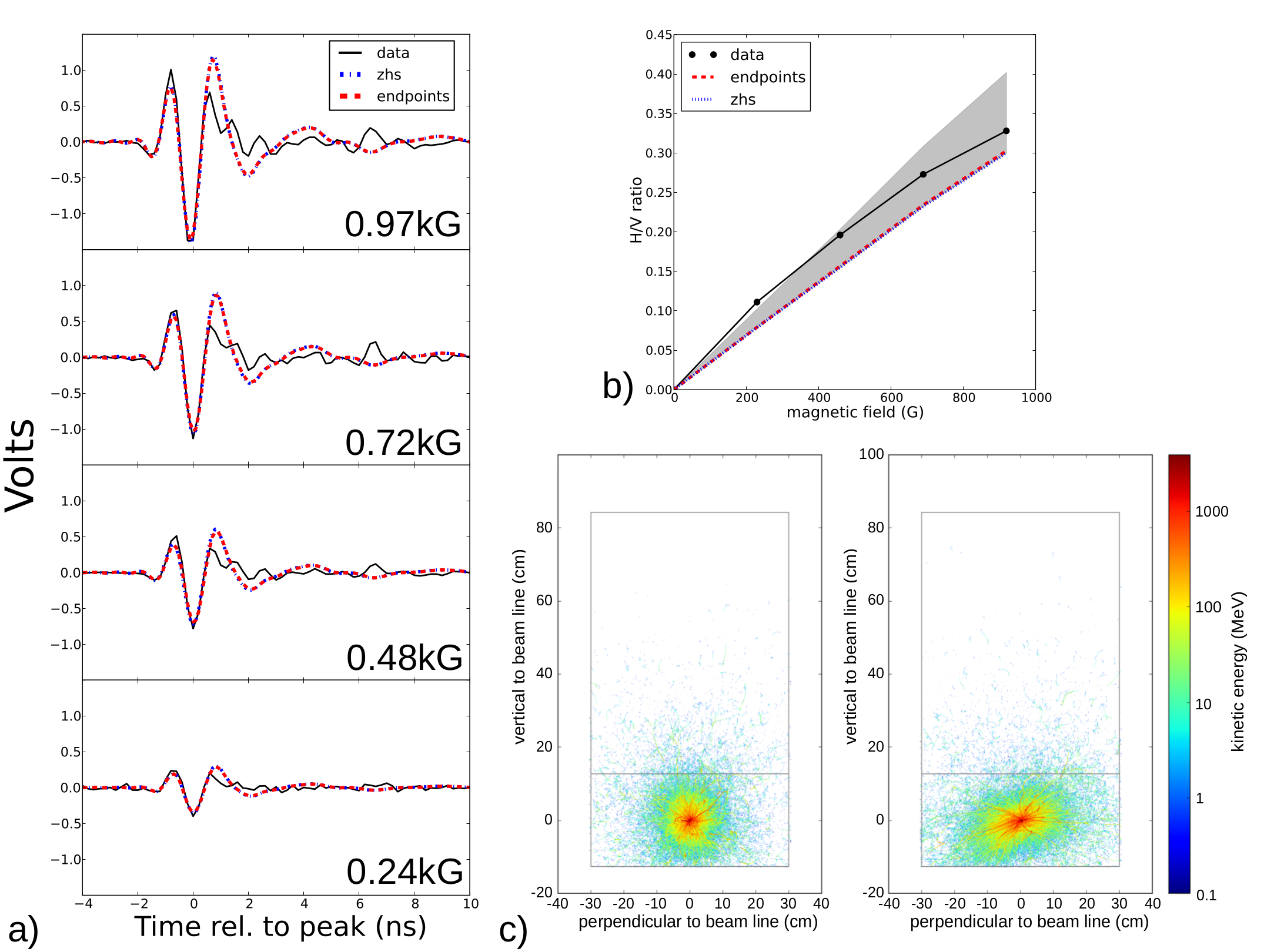}
\caption{a) Measurements of the scaling of the peak amplitude of the magnetically induced component of the radio signal with increasing magnetic field strength. b) Measured and simulated ratio of the peak amplitudes of the magnetically induced component and the component induced by the charge excess during the shower development. The gray band on top of the simulations indicates the possible impact of the reflections. c) Distribution of particles in the shower for different magnetic field strengths (left: $B=0\,\mbox{G}$, right: $B=1940\,\mbox{G}$) looking along the beam line. The color code represents the kinetic energy of the particles.}
\figlab{Scaling}       % Give a unique label
\end{figure}
To study the linearity of the scaling, the ratio of the peak amplitudes of the magnetically induced component to that induced by the charge excess during the shower development is shown in \figref{Scaling}b. As expected the ratio rises with increasing magnetic field strength. 
However, we observe a flattening of the ratio towards higher magnetic field strengths and a power law fit to the data does not yield an exponent of $1$, meaning that the scaling is not linear. 
The simulations seem not to reproduce the flattening for the studied magnetic field strength. But the uncertainty introduced by internal reflection, indicated by a gray band in the plot, contains the measured ratios and even indicates a slight flattening for the higher magnetic field strengths.

There are different possibilities to explain the measured flattening of the ratio of the peak amplitudes of the magnetically induced component with respect to the component induced by the charge excess during the shower development: 
First, there could be a so far unknown physical effect, like a saturation effect in the drift velocity of the electrons and positrons, which is not reproduced in the simulations. 
Secondly, to exclude the possibility that high-energy particles escape the target and therefore don not contribute to the radio signal of the shower, the distribution of particles in the shower without any magnetic field is studied (see \figref{Scaling}). As shown by the color code, no high-energy particles leave the target. 
But, the particle distribution with an induced magnetic field strength of an exaggerated value of $B=1940\,\mbox{G}$ shows that the particle shower is rotated out of the horizontal alignment due to the non-uniform three-dimensional magnetic field. 
This leads to another possible explanation namely that the geometry of the radio emission, assuming that the magnetically induced radio component is aligned with the horizontal channel of the antenna and the charge excess induced component is aligned with the vertical component, is not valid for high magnetic fields. A rotation of the particle showers would also rotate the electric field vector of the magnetically induced component and thus the Askaryan and magnetic component of the radio emission can not be measured independently in the T-510 experiment. This hypothesis will be tested by a more detailed simulation study.

\section{Summary}
\vspace{-0.15cm}
The SLAC T-510 experiments constitutes the first laboratory benchmark of radio-frequency radiation from electromagnetic cascades under the influence of a magnetic field~\cite{PRL}.
The experiment confirmed that simulations based on the first-principle formalisms of electrodynamics can predict the absolute scale of the radio emission accurately to within the systematic uncertainty of the measurements. Internal reflections at the bottom surface of the target are the main limiting uncertainty in this experiment.
The magnetically induced radio signal component shows a near-linear scaling with rising magnetic field strength, but a flattening towards higher magnetic field strengths is measured. There are  indications that the flattening can be explained by a rotation of the particle shower due the non-uniform three-dimensional magnetic field. A future more-detailed simulation study can answer this question.

%
% BibTeX or Biber users please use (the style is already called in the class, ensure that the "woc.bst" style is in your local directory)
% \bibliography{name or your bibliography database}

\begin{thebibliography}{}

\bibitem{AIRES}
S.~J.~Sciutto, arXiv:astro-ph/9911331.
%, {\it AIRES: A system for air-shower simulations (Version 2.2.0)}
%
\bibitem{CORSIKA}
D.~Heck, {\it The Air Shower Simulation Program CORSIKA}, https://web.ikp.kit.edu/corsika/.
%
\bibitem{Endpoint}
C.~W.~James, H.~Falcke, T.~Huege and M.~Ludwig, arXiv:1007.4146 [physics.class-ph].
%, {\it General description of electromagnetic radiation processes based on instantaneous charge acceleration in `endpoints'}
%
\bibitem{REAS3}
% M.~Ludwig, T.~Huege, O.~Scholten, K.~D.~de Vries, arXiv:1202.3352 [astro-ph.HE].
M.~Ludwig, T.~Huege, Astroparticle Physics 34, 438-446, 2011.
%
\bibitem{CoREAS}
T.~Huege, M.~Ludwig, C.~W.~James, arXiv:1301.2132 [astro-ph.HE].
%, {\it Simulating radio emission from air-showers with CoREAS}
%
\bibitem{ZHS}
J.~Alvarez-Muniz, A.~ Romero-Wolf, E.~ Zas, arXiv:1002.3873v1 [astro-ph.HE].
%, {\it Cherenkov radio pulses from electromagnetic showers in the time-domain}
%
\bibitem{ZHAireS}
J.~Alvarez-Muniz, W.~R.~Carvalho Jr., E.~Zas, arXiv:1107.1189 [astro-ph.HE].
%, {\it Monte-Carlo simulations of radio pulses in atmospheric showers using ZHAireS}
% 
\bibitem{Katie}
K.~Mulrey, et al., \textit{SLAC T-510: Accelerator measurements of radio emission from particle cascades in the presence of a magnetic field}, these preceedings.
%
\bibitem{AnneICRC}
A.~Zilles et al., PoS(ICRC2015)313.
%
\bibitem{PRL}
K.~Belov et al., Phys. Rev. Lett. 116, 141103 2016.


\end{thebibliography}

\end{document}